\documentclass[a4paper,10pt]{article} 
\usepackage[margin=0pt]{geometry}

\usepackage{pgfplots}
\pgfplotsset{compat=1.15}
\usepackage{mathrsfs}
\usetikzlibrary{arrows}

\usepackage{pstricks,pst-node}

\usetikzlibrary{shapes.misc}
\tikzset{cross/.style={cross out, draw=black, fill=none, minimum size=2*(#1-\pgflinewidth), inner sep=0pt, outer sep=0pt}, cross/.default={2pt}}

\usepackage[T1]{fontenc}

\usepackage[]{amsmath, amssymb, amsthm, tabularx}

\usepackage[]{a4, xcolor, here}

\usepackage[]{pstricks, pst-text, pst-node, pst-tree, gastex}

\usepackage{tikz}

\usepackage[tikz]{bclogo}

\usepackage{comment}

\usepackage{boites,graphicx}

\usepackage{soul}

\definecolor{pastelyellow}{rgb}{0.99, 0.99, 0.59}
\definecolor{aqua}{rgb}{0.0, 1.0, 1.0}
\definecolor{aquamarine}{rgb}{0.5, 1.0, 0.83}
\definecolor{bananayellow}{rgb}{1.0, 0.88, 0.21}
\definecolor{burgundy}{rgb}{0.5, 0.0, 0.13}
\definecolor{ao(english)}{rgb}{0.0, 0.5, 0.0}

\setul{}{0.2ex}
\setulcolor{bananayellow}

\usepackage[normalem]{ulem}

\usepackage{mathrsfs}

\usepackage{stmaryrd}

\usepackage{enumerate}

\usepackage{paralist}

\usepackage{multienum}

\usepackage{multicol}

\usepackage{fancyhdr}

\usepackage{datetime}

\usepackage[contents={},opacity=1,scale=1.6,
color=gray!90]{background}
\usepackage{ifthen}

\usepackage[]{hyperref}

\hypersetup{
colorlinks = true,
linkcolor = red,
anchorcolor = black,
citecolor = blue, 
filecolor = cyan,
menucolor = red,
runcolor = cyan,
urlcolor = blue,
linkbordercolor = {white},
linktocpage = true
}

\newtheorem{theorem}{Theorem}[section]
\newtheorem{proposition}[theorem]{Proposition}
\newtheorem{lemma}[theorem]{Lemma}
\newtheorem{corollary}[theorem]{Corollary}

\theoremstyle{definition}
\newtheorem{definition}[theorem]{Definition}

\newtheorem{example}[theorem]{Example}
\newtheorem{remark}[theorem]{Remark}

\makeatletter
\def\thmhead@plain#1#2#3{%
\thmname{#1}\thmnumber{\@ifnotempty{#1}{ }\@upn{#2}}%
\thmnote{ {\the\thm@notefont#3}}}
\let\thmhead\thmhead@plain
\makeatother

\newcommand{\bt}{\mathbf{t}}
\newcommand{\bd}{\mathbf{d}}

\newcommand{\cC}{\mathcal{C}}

\newcommand{\cF}{\mathcal{F}}
\newcommand{\cG}{\mathcal{G}}

\newcommand{\cU}{\mathcal{U}}
\newcommand{\cV}{\mathcal{V}}

\newcommand{\lcm}{\mathrm{lcm}}
\newcommand{\stab}{\mathrm{Stab}}

\newcommand{\orb}{\mathrm{Orb}}

\newcommand{\bbZ}{{\mathbb Z}}
\newcommand{\bbF}{{\mathbb F}}

\renewcommand{\geq}{\geqslant}
\renewcommand{\leq}{\leqslant}

\newcommand{\tipon}{{(\bt, n)}}

\begin{document}

\renewcommand{\headrulewidth}{0pt}

\rhead{ }
\chead{\scriptsize A new invariant for cyclic orbit flag codes}
\lhead{ }

\title{ A new invariant for cyclic orbit flag codes 
\renewcommand\thefootnote{\arabic{footnote}}\footnotemark[1]
}

\author{\renewcommand\thefootnote{\arabic{footnote}}
Clementa Alonso-Gonz\'alez\footnotemark[2] \ and \  Miguel \'Angel Navarro-P\'erez\footnotemark[3]}

\footnotetext[1]{The authors receive financial support from Ministerio de Ciencia e Innovaci\'on (PID2019-108668GB-I00).}

\footnotetext[2]{Departamento de Matem\'aticas, Universidad de Alicante, Carr. de San Vicente del Raspeig, s/n, 03690, San Vicente del Raspeig, Alicante (Spain).}

\footnotetext[3]{Departamento \ de Matem\'aticas, Universidad Carlos III de Madrid, Avda. de la Universidad, 30, 28911, Leganés, Madrid (Spain).\\
Contact: M. A. Navarro-P\'erez. \ Email: \href{mailto:mignavar@math.uc3m.es}{mignavar@math.uc3m.es}}


\maketitle

\begin{abstract}
In the network coding framework, given a prime power $q$ and the vector space $\mathbb{F}_q^n$, a constant type flag code is a set of nested sequences of $\mathbb{F}_q$-subspaces (flags) with the same increasing sequence of dimensions (the type of the flag). If a flag code arises as the orbit under the action of a cyclic subgroup of the general linear group over a flag, we say that it is a \textit{cyclic orbit flag code}. Among the parameters of such a family of codes, we have its \textit{best friend}, that is the largest field over which all the subspaces in the generating flag are vector spaces.  This object permits to compute the cardinality of the code and estimate its minimum distance. 
However, as it occurs with other absolute parameters of a flag code, the information given by the best friend is not complete in many cases due to the fact that it can be obtained in different ways. In this work, we present a new invariant, the \textit{best friend vector}, that captures the specific way the best friend can be unfolded. Furthermore, throughout the paper we analyze the strong underlying interaction between this invariant and other parameters such as the cardinality, the flag distance, or the type vector, and how it conditions them. Finally, we investigate the realizability of a prescribed best friend vector in a vector space.

\end{abstract}



\section{Introduction}\label{sec:Introduction}

 Network coding was presented as an effective method to transmit information encoded as vector spaces over a finite field (see \cite{NetworkCoding08}). The use of flags in the context of network coding was first introduced in \cite{LiebNebeVaz18}. Fixed a prime power $q$ we can take the field extension $\mathbb{F}_{q^n}$, with $n \geq 2$ a positive integer, and consider it as a vector space of dimension $n$ over $\mathbb{F}_q$.  A \textit{flag} on $\mathbb{F}_{q^n}$ is a sequence of nested proper subspaces of $\mathbb{F}_{q^n}$. The increasing sequence of dimensions of the subspaces in a flag is called its \textit{type}. Collections of flags of constant type are denominated \emph{flag codes}. The recent works \cite{CasoPlanarOrb,Cotas, CasoPlanar, Kurz20}, among others, show a growing interest in this subject.

Flag codes can be seen as a generalization of \textit{constant dimension codes}, sets of subspaces of $\bbF_{q^n}$ sharing the same dimension (for more information on this family of codes, consult \cite{TrautRosen18} and the references therein). A special family of constant dimension codes is the one of \emph{orbit (subspace) codes}, introduced in \cite{TrautManRos10}, as orbits under the action of a subgroup of the general linear group on the set of subspaces of $\bbF_{q^n}$. In particular, in \cite{TrautManBraunRos13}, the authors focused on the situation in which the acting group is cyclic and define \emph{cyclic orbit (subspace) codes}.

In \cite{GLMorTro15}, Gluesing-Luerssen \emph{et al.} studied cyclic orbit codes under the natural action of multiplicative subgroups of $\bbF_{q^n}^\ast$ (cyclic groups as well) on $\bbF_q$-vector spaces of $\bbF_{q^n}$. Following this approach, in \cite{Cyclic}, the authors defined \emph{cyclic orbit flag codes} in the same way but considering the action of such subgroups on flags. Also in  \cite{GLMorTro15}, it was introduced the notion of \emph{best friend} of a cyclic orbit code as the main tool for the study of this family of codes: fixed a generating subspace $\cU$  and its corresponding orbit $\orb(\cU)$, the \emph{best friend} of $\cU$, and then of $\orb(U)$, is the largest subfield of $\bbF_{q^n}$ over which $\cU$ is a vector space. In \cite{Cyclic}, this concept was generalized to the flag codes framework by defining the \emph{best friend} of a flag code of the form $\orb(\cF)$ as the largest subfield of $\bbF_{q^n}$ over which every subspace in the generating flag $\cF$ is a vector space. Similarly to the case of constant dimension codes, the knowledge of the best friend of a cyclic orbit flag code $\orb(\cF)$ permits to determine directly its size and to give estimates for its distance. 

Nevertheless, as it happens with other parameters of a flag code, the information provided by the best friend associated to  $\orb(\cF)$, could not be sufficient to determine specific properties of the code, which makes necessary to  take into account how these parameters are unfolded according to the nested structure of the flag. In the case of the flag distance, this viewpoint was developed in \cite{Cotas} where the authors coined the concept of \emph{distance vector} associated with a pair of flags to describe how a flag distance value is obtained as the sum of subspace distances between the corresponding subspaces. The knowledge of the distance vectors set associated with the minimum distance of a flag code provides more precise information that allows us to derive important properties (see \cite{Cotas}). In the paper at hand, we propose a new invariant, the \textit{best friend vector} of a cyclic orbit flag code, that describes how the best friend is obtained and, consequently, encloses more accurate information than it. In fact, throughout this article it will be revealed the strong underlying interplay between the best friend vector and other invariants such as the cardinality, the flag distance and the type vector, and how them are conditioned by this new object.

The text is organized as follows. In Section \ref{sec:Preliminaries}, we recall the basics on constant dimension codes, cyclic orbit (subspace) codes and some known ideas related to flag codes. In Section \ref{sec: BF and BF vector}, we focus on cyclic orbit flag codes, putting the accent on the concept of best friend of a flag. Then we introduce the notion of best friend vector and explore some features of this new invariant of a flag (and, consequently, of the cyclic orbit flag code that it generates). Section \ref{sec:interplay} is devoted to study how the best friend vector of a flag influences the rest of parameters of the code. More precisely, in Subsection \ref{subsec: bf and dist}, we derive new lower and upper bounds for the minimum distance of $\orb(\cF)$ in terms of the best friend vector of $\cF$ and see that they considerably improve those obtained in \cite{Cyclic, GenGalois} just taking into consideration the best friend of the flag. Later, in Subsection \ref{subsec: bf and type}, we observe that having a prescribed best friend vector is not always compatible with a given type vector or a given value of $n$  and investigate the interaction between these parameters. This study is carried out in two steps: first we consider best friend vectors of length $r=2$ for which we present a characterization of those that are realizable in $\bbF_{q^n}$. Secondly, we address the case of length $r>2$ and provide a sufficient condition on $n$ for the realizability of a prescribed best friend vector by developing a systematic construction of appropriate flags. Finally, we propose a reciprocal to our result for some special best friend vector choices. Last, in Section \ref{sec: conclusions}, we summarize the advances provided in our work and present some related open questions.

\section{Preliminaries}\label{sec:Preliminaries}

Let $q$ be a prime power and $\bbF_q$ the finite field with $q$ elements. For every integer $n\geq 2$, we write $\bbF_{q^n}$ to denote the extension field with $q^n$ elements, which also is an $n$-dimensional vector space over $\bbF_q$. For every $1\leq k < n,$ \emph{the Grassmannian (of dimension $k$ of $\bbF_{q^n}$)} is the set $\cG_q(k, n)$ of $\bbF_q$-subspaces of dimension $k$ of $\bbF_{q^n}$. This set is a metric space endowed with the \emph{subspace distance}, computed as 
\begin{equation}\label{eq: subspace distance}
	d_S(\cU, \cV)= \dim(\cU + \cV) - \dim(\cU \cap \cV) = 2 ( k -\dim(\cU \cap \cV))\leq \min\{ 2k, 2(n-k)\}.
\end{equation}

A \emph{constant dimension code} of $\bbF_{q^n}$ is a nonempty subset $\cC\subseteq\cG_q(k, n)$ and its \emph{minimum distance} is the value 
$$
d_S(\cC)=\min\{ d_S(\cU, \cV) \ | \ \cU, \cV\in\cC, \ \cU\neq \cV\}
$$
whenever $|\cC|\geq 2$. According to (\ref{eq: subspace distance}), it is an even integer $0\leq d_S(\cC)\leq \min\{ 2k, 2(n-k)\}.$ In case that $|\cC|=1,$ we simply put $d_S(\cC)=0.$ 

The group $\bbF_{q^n}^\ast$ acts naturally on the Grassmannian in this way: $\cU  \alpha = \{ u\alpha \ | \ u\in\cU\},$ for every $\cU\in\cG_q(k, n)$ and every $\alpha\in\bbF_{q^n}^\ast$. In \cite{GLMorTro15}, the authors use this action and consider constant dimension codes arising as its orbits.
\begin{definition}
	Given $\cU\in\cG(k, n)$, the set
	$$
	\orb(\cU) = \{ \cU \alpha \ | \ \alpha\in \bbF_{q^n}^\ast\} \subseteq\cG_q(k, n)
	$$
	is a constant dimension code called the \emph{cyclic orbit code generated by $\cU$}. The stabilizer subgroup of $\cU$ in $\bbF_{q^n}^\ast$ is 
	$
	\stab(\cU) = \{ \alpha\in\bbF_{q^n}^\ast \ | \ \cU\alpha=\cU\}
	$
	and it holds $|\orb(\cU)| =\frac{|\bbF_{q^n}^\ast|}{|\stab(\cU)|}.$
\end{definition}

In \cite{GLMorTro15}, it was also introduced the notion of \emph{best friend} of a subspace.
\begin{definition}\label{def: bf subspace}
	Let $\cU$ be a subspace of $\bbF_{q^n}$, a subfield $\bbF_{q^m}$ of $\bbF_{q^n}$ is a \emph{friend} of $\cU$ if $\cU$ is a vector space over $\bbF_{q^m}$. The largest subfield satisfying this property is called the \emph{best friend} of $\cU$.
\end{definition}

The knowledge of the best friend of a subspace gives information about the parameters and features of the cyclic orbit code that it generates. As we can see in the following results, it describes some properties of the minimum distance and also determines the stabilizer subgroup (then the cardinality of the code). 

\begin{theorem}\label{th: bf and dist}(\cite[Lemma 4.1]{GLMorTro15})
	Let $\cU\in\cG_q(k, n),$ and assume that $\bbF_{q^m}$ is a friend of $\cU$. Then $m$ is a divisor of $\gcd(k, n)$ and $2m$ divides $d_S(\cU, \cU\alpha)$, for every $\alpha\in\bbF_{q^n}^\ast.$ In particular, $2m$ divides $d_S(\orb(\cU)).$
\end{theorem}

\begin{theorem}\label{th: bf and size subspaces}(\cite[Cor. 3.13]{GLMorTro15})
	Let $\cU$ be a subspace of $\bbF_{q^n}$. The following statements are equivalent:
	\begin{enumerate}
		\item The orbit $\orb(\cU)$ contains $\frac{q^n-1}{q^m-1}$ elements, 
		\item the subfield $\bbF_{q^m}$ is the best friend of $\cU$ and
		\item $\stab(\cU)=\bbF_{q^m}^\ast.$
	\end{enumerate}
\end{theorem}

In \cite{{LiebNebeVaz18}}, the authors introduce the family of flag codes as a generalization of constant dimension codes. In this new setting, codewords are flags defined as follows.
\begin{definition}\label{def: flag}
	Given integers $0 < t_1 < \dots < t_r < n$, a sequence $\cF = (\cF_1, \dots, \cF_r)$
	is called \emph{flag of type $(t_1, \dots, t_r)$} on  $\bbF_{q^n}$ if 
	$$
	\cF_1\subsetneq \dots \subsetneq \cF_r \subsetneq \bbF_{q^n}
	$$
	and $\cF_i\in\cG_q(t_i, n),$ for every $1\leq i\leq r$. 
\end{definition}

The set of flags of a given type on $\bbF_{q^n}$ is also a metric space.
Given two flags $\cF, \cF'$ of type $(t_1, \dots, t_r)$ on $\bbF_{q^n}$, their \emph{flag distance} is
\begin{equation}
	d_f(\cF, \cF')=\sum_{i=1}^r d_S(\cF_i, \cF'_i).
\end{equation}
According to the definition of the subspace distance and expression (\ref{eq: subspace distance}), $d_f(\cF, \cF')$ is an even integer satisfying 
$
0\leq d_f(\cF, \cF') \leq D^\tipon,
$
where $D^\tipon$ is the maximum possible value of the flag distance for type $(t_1, \dots, t_n)$, that is, 
\begin{equation}\label{eq: max dist}
	D^\tipon = 2\left( \sum_{t_i\leq \frac{n}{2}} t_i + \sum_{t_i > \frac{n}{2}} (n-t_i) \right).
\end{equation}

\noindent In this new framework, flag codes are defined as follows.
\begin{definition}
	A \emph{flag code} $\cC$ of type $(t_1, \dots, t_r)$ on $\bbF_{q^n}$ is a nonempty set of flags of this type. Its \emph{minimum distance} is the value
	$$
	d_f(\cC)=\min\{d_f(\cF, \cF') \ | \ \cF, \cF'\in\cC, \ \cF\neq \cF'\}
	$$
	if $|\cC|\geq 2.$ If $|\cC|=1,$ we put $d_f(\cC)=0$.
\end{definition}

There is a family of constant dimension codes naturally induced by a flag code, introduced for the first time in \cite{CasoPlanar}.
\begin{definition}
	Let $\cC$ be a flag code of type $(t_1, \dots, t_r)$ on $\bbF_{q^n}$. For every $1\leq i\leq r$, the \emph{$i$-th projected code} of $\cC$ is  the constant dimension code
	$$
	\cC_i=\{ \cF_i \ | \ \cF\in\cC\}\subseteq \cG_q(t_i, n).
	$$
\end{definition}

In the same paper, the authors introduce the family of flag codes attaining the maximum possible distance (see (\ref{eq: max dist})), they are called \emph{optimum distance flag codes}, and characterize them in terms of the projected codes. 
\begin{theorem}
	A flag code $\cC$ of type $(t_1, \dots, t_r)$ on $\bbF_{q^n}$ is an optimum distance flag code, i.e., $d_f(\cC)=D^\tipon$, if, and only if, the following statements hold:
	\begin{enumerate}
		\item $d_s(\cC_i)=\min\{2t_i, 2(n-t_i)\},$ for every $1\leq i\leq r$ and
		\item $|\cC|=|\cC_1|=\dots=|\cC_r|.$
	\end{enumerate}
\end{theorem}
Notice that, in this result, we can already appreciate a first connection between the parameters (minimum distance and size) of flag codes: attaining the maximum possible distance requires certain conditions on the cardinality of the flag code.

For other values of the minimum distance, characterizing flag codes in terms of their projected codes is still an open problem. This is due to the fact that the maximum possible distance value $D^\tipon$ can only be obtained by summing the maximum possible subspace distances for every dimension in the type vector. Out of this case, lower values of the flag distance can be reached from the sum of different combinations of subspace distances. To deal with this problem, in \cite{Cotas} the authors introduce the concept of \emph{distance vector} associated with a pair of flags as follows:
$$
\bd(\cF, \cF)=(d_S(\cF_1, \cF'_1), \dots, d_S(\cF_r, \cF'_r))\in 2\bbZ^r.
$$
Notice that the sum of the components of $\bd(\cF, \cF')$ is the value $d_f(\cF, \cF')$.
Similarly, a vector $\bd\in 2\bbZ^r$ is called a \emph{distance vector associated to a distance value $0\leq d\leq D^\tipon$} if there exist flags $\cF, \cF'$ (of type $(t_1, \dots, t_r)$ on $\bbF_{q^n}$) such that $\bd=\bd(\cF, \cF')$ and $d=d_f(\cF, \cF').$  In the same paper, the next characterization of distance vectors is given.
\begin{theorem}\label{th: char dist vect} (\cite[Th. 3.9]{Cotas})
	Consider an even integer $0\leq d\leq D^\tipon.$ A vector $\bd=(d_1 ,\dots, d_r) \in 2\bbZ^r$ is a distance vector associated to $d$ if, and only if, the following statements hold.
	\begin{enumerate}
		\item $\sum_{i=1}^r d_i=d$,
		\item $0\leq d_i\leq \min\{2t_i, 2(n-t_i)\},$ for every $1\leq i\leq r$ and
		\item $|d_{i+1}-d_i|\leq 2(t_{i+1}-t_i),$ for every $1\leq i\leq r-1.$
	\end{enumerate}
\end{theorem}
The concept of distance vector has been extremely useful to better understand the different possible combinations that can provide the same distance value and to provide upper bounds for the cardinality of flag codes having a prescribed minimum distance for every choice of the type vector. This fact demonstrates again the intrinsic connection that exists between the parameters (minimum distance and size) of a flag code. Also in \cite{Cotas}, the authors use other particular values of the flag distance that will be also useful in the paper at hand. 

\begin{definition}
	Consider the type vector $(t_1, \dots, t_r)$ on $\bbF_{q^n}$ and integers $1\leq M\leq r$, $1\leq i_1 < \dots < i_M \leq r.$ We write $D(i_1, \dots, i_M)^\tipon$ to denote the maximum possible distance that can be attained with a distance vector having zeroes at the positions $i_1, \dots, i_M.$ In other words
	$$
	D^\tipon(i_1, \dots, i_M)= \max \{d_f(\cF, \cF') \ | \ \cF_{i_j}=\cF'_{i_j}, \ 1\leq j\leq M\}.
	$$
\end{definition}

According to Theorem \ref{th: char dist vect}, the value $D^\tipon(i_1, \dots, i_M)$ can be computed explicitly and satisfies:

\begin{equation}
	D^\tipon(i_1, \dots, i_M) = \sum_{k=1}^r \underset{1\leq j\leq M}{\min}\{ 2t_k, \ 2(n-t_k), \ 2|t_k-t_{i_j}| \}.
\end{equation}

As it occurs for subspaces of $\bbF_{q^n}$, the multiplication by nonzero elements in $\bbF_{q^n}$ defines an action on the set of flags. More precisely, given a flag $\cF=(\cF_1, \dots, \cF_r)$ of type $(t_1, \dots, t_r)$ on $\bbF_{q^n}$ and $\alpha\in\bbF_{q^m}^\ast$, we have that $\cF\alpha = (\cF_1\alpha, \dots, \cF_r\alpha)
$ is a flag of the same type.

\begin{definition}
	Let $\cF=(\cF_1, \dots, \cF_r)$ of type $(t_1, \dots, t_r)$ on $\bbF_{q^n}$. The set
	$$
	\orb(\cF)=\{\cF\alpha \ | \ \alpha\in\bbF_{q^n}^\ast\}.
	$$
	is called the \emph{cyclic orbit flag code generated by $\cF$} and $\stab(\cF)=\{ \alpha\in\bbF_{q^n}^\ast \ | \ \cF=\cF\alpha\}$ is its stabilizer subgroup under the action of $\bbF_{q^n}^\ast.$
\end{definition}

As for cyclic orbit subspace codes, we can define the concept of best friend of a cyclic orbit flag code $\orb(\cF)$. It was introduced in \cite{Cyclic} and used also to obtain certain information about its parameters. In the following section we will recall this notion and how the parameters of cyclic orbit flag codes are influenced by it. Moreover, as it happens with other absolute parameters as the flag distance, the information given by the best friend is not complete in many cases. If the concept of distance vector comes to help in the determination of properties of a flag code beyond those derived from the flag distance value, in this paper, we present the new concept of  \emph{best friend vector} of a flag in order to obtain more precise information about cyclic orbit flag codes. Furthermore, the use of this new invariant allows us to evince the strong underlying interaction between different parameters.

\section{Best friend and best friend vector of a flag}\label{sec: BF and BF vector}

Following the viewpoint developed in \cite{GLMorTro15} for the case of cyclic orbit codes, in  \cite{Cyclic}, the authors introduce the \textit{best friend }of a flag $\cF$. With the help of this new object they can compute the cardinality and estimate the distance of the cyclic orbit flag code generated by $\cF$. However, there are some properties of $\orb(\cF)$ that are not completely determined from its best friend but rather by the way it is obtained. In this section we go further and propose a finer invariant associated with $\cF$, its \textit{best friend vector}, and use it to obtain more precise information about $\orb(\cF)$, its parameters and those of its projected codes, and how they are related. 

\subsection{Best friend of a flag}

Le us recall the definition of best friend of a flag given in  \cite{Cyclic}.

\begin{definition}\label{def: bf flag}
Given a flag $\cF=(\cF_1, \dots, \cF_r)$ on $\bbF_{q^n}$, we say that a subfield $\bbF_{q^m}$ of $\bbF_{q^n}$ is \emph{a friend} of $\cF$ if every $\cF_i$ is a vector space over $\bbF_{q^m}$. Among the friends of $\cF$, the largest one is called its \emph{best friend}.  
\end{definition}

\begin{remark}\label{rem: 1 in F1}
Notice that the vector space structure of any subspace of $\bbF_{q^n}$ is preserved under multiplication by elements in $\bbF_{q^n}^\ast$. As a consequence, given a flag $\cF$ on $\bbF_{q^n}$, it is clear that every flag in the orbit $\orb(\cF)$ has exactly the same best friend. Thus, we will speak indistinctly about the best friend of a flag $\cF$ or the best friend of the code $\orb(\cF).$
Moreover, without loss of generality, in some cases, we will consider flags $\cF=(\cF_1, \dots, \cF_r)$ on $\bbF_{q^n}$ such that $1\in\cF_1$. This is not restrictive at all since, if for a given flag $\cF$ we take $\alpha\in\cF_1\setminus\{0\}$, we have that $\orb(\cF)=\orb(\cF\alpha^{-1})$ and, at the same time, it holds $1\in\cF_1\alpha^{-1}$. Under the assumption $1\in\cF_1$, every subspace in the flag contains its best friend.
\end{remark}

As it happens for cyclic orbit subspace codes, the best friend of a flag $\cF$ is clearly connected with the orbit size of $\orb(\cF)$. 
\begin{theorem}\label{th: bf and size flag}(\cite[Prop. 4.1]{Cyclic})
The size of $\orb(\cF)$ is $\frac{q^n-1}{q^m-1}$ for some divisor $m$ of $n$ if, and only if, $\bbF_{q^m}$ is the best friend of $\cF$.
\end{theorem}

On the other hand, if $\bbF_{q^m}$ is the best friend of a flag  $\cF$, then  $\cF$ if a flag of type $(ms_1, \ldots, ms_r),$ for some integers $1\leq s_1 < \dots < s_r < s = \frac{n}{m}$, and we can provide some bounds for the distance of $\orb(\cF)$.

	\begin{proposition}\label{prop: distance bounds BF}
	Let $\cF$ be a flag of type $(ms_1, \ldots, ms_r)$ on $\bbF_{q^ms}$ with the subfield $\bbF_{q^m}$ as its best friend and take $\beta\in\bbF_{q^n}^\ast.$  Then  $2m$ divides $d_f(\orb(\cF))$ and it holds
	\begin{equation}\label{eq: distance bounds}
		2m \leq d_f(\orb(\cF))  \leq  2m \left( \sum_{s_i \leq \lfloor \frac{s}{2}\rfloor} s_i + \sum_{s_i > \lfloor \frac{s}{2}\rfloor} (s-s_i) \right) =  D^{((ms_1, \dots, ms_r), ms)}.  
	\end{equation}
\end{proposition}

Notice that, according to Definition \ref{def: bf flag}, if $\bbF_{q^m}$ is a friend of a flag $\cF =(\cF_1, \dots, \cF_r)$, then it is a friend of all the subspaces $\cF_i$.  Even more, to compute the best friend of a flag it is enough to know the ones of its subspaces. In \cite{Cyclic}, the next property is proved.

\begin{proposition}\label{prop: bf intersection bfs}(\cite[Cor. 3.18]{Cyclic})
	The best friend of a flag is the intersection of the best friends of its subspaces.
\end{proposition}

A special case of cyclic orbit flag codes is the one of Galois flag codes proposed in \cite{Cyclic}. 
\begin{example}
	Given a sequence of divisors $t_1, \dots, t_r$ of $n$ such that every $t_i$ divides $t_{i+1}$, the \emph{Galois flag} of type $(t_1, \dots, t_r)$ is the sequence of nested subfields
	\begin{equation}
		\cF=(\bbF_{q^{t_1}}, \dots, \bbF_{q^{t_r}})
	\end{equation}
	of $\bbF_{q^n}.$ The code $\orb(\cF)$ is called \emph{the Galois flag code} of type  $(t_1, \dots, t_r)$. Here, each subspace $\cF_i$ is its own best friend and, in particular, the best friend of $\cF$ is its first subspace $\bbF_{q^{t_1}}$.
\end{example}

From Proposition \ref{prop: bf intersection bfs} it is immediate to realize that we can have two flags $\cF, \cF'$ of type $(t_1, \ldots, t_r)$ that share the same best friend but such that some of their corresponding respective subspaces $\cF_i, \cF_i'$ do not satisfy this condition. 

\begin{example}\label{ex:flags different best friend vector} For type $(2,4,8)$ consider the Galois flag $\cF=(\bbF_{q^2}, \bbF_{q^4}, \bbF_{q^8})$ and the flag $\cF'=(\bbF_{q^2}, \bbF_{q^2}\oplus \bbF_{q^2}\beta, \bbF_{q^8})$ with $\beta \in \bbF_{q^8}\setminus\bbF_{q^4}$.  Notice that $\cF'_2$ is an $\bbF_{q^2}$-vector space with $1\in\cF'_2$ and $\dim_q(\cF'_2)=4$. Moreover, it is different from $\bbF_{q^4}$ by the choice of $\beta$. Then its best friend is clearly $\bbF_{q^2}$. As a consequence, the best friend of both $\cF$  and $\cF'$ is $\bbF_{q^2}$, whereas subspaces $\cF_2$ and $\cF'_2$ have the subfields $\bbF_{q^4}$ and $\bbF_{q^2}$ as their best friends, respectively.
\end{example} 

Clearly, in light of Theorem \ref{th: bf and size flag} and Proposition \ref{prop: distance bounds BF}, for the flags $\cF, \cF'$ of the previous example, the codes $\orb(\cF)$ and $\orb(\cF')$ have the same cardinality and the same estimates for the minimum distance. Nevertheless, we wonder if the fact that their best friend is equal but obtained in different ways, provokes that other of their parameters or properties might be different. To this end, in the following subsection we introduce the notion of \textit{best friend vector} of a flag. In Section \ref{sec:interplay} we discuss the relationship of this new invariant with other parameters of a cyclic orbit flag codes such as the distance and the type vector.

\subsection{Best friend vector of a flag}

The best friend vector of a flag $\cF$ specifies the sequence of best friends of its subspaces, that is, the way we obtain the best friend of $\cF$. More in precise:

\begin{definition}\label{def: bf vector flag}
	Consider a flag $\cF=(\cF_1, \dots, \cF_r)$ on $\bbF_{q^n}$ such that $\bbF_{q^{m_i}}$ is the best friend of $\cF_i$, for any $i \in \{1, \ldots, r\}$. Then the sequence $(m_1, \ldots, m_r)$ will be called the \emph{best friend vector of $\cF$}.
\end{definition}

As a consequence of Proposition \ref{prop: bf intersection bfs} and the previous definition, the next result clearly holds.
\begin{proposition}
Let $\cF$ be a flag of type $(t_1, \dots, t_r)$ on $\bbF_{q^n}$ with best friend vector $(m_1, \ldots, m_r)$. Then $m_i$ divides $t_i$ for every $1\leq i\leq r$. Moreover, if $m=\gcd(m_1, \dots, m_r)$, then the subfield $\bbF_{q^m}$ is the best friend of $\cF$.
\end{proposition}

As underlined in Remark \ref{rem: 1 in F1}, given a flag $\cF=(\cF_1, \ldots, \cF_r)$ on $\bbF_{q^n}$, we have that every subspace in the orbit $\orb(\cF_i)$ has exactly the same best friend. Hence every flag in $\orb(\cF)$ share the same best friend vector.  Thus, also in this case, we speak indistinctly about the best friend vector of a flag $\cF$ or the best friend vector of the code $\orb(\cF).$

At this point, a natural question is if a best friend vector must satisfy any kind of property beyond the fact that their entries must be divisors of $n$. Example \ref{ex:flags different best friend vector} shows that, even if the subspaces in a flag are nested, that is, the entries in the type vector of a flag give a strictly increasing sequence of dimensions, this property is not transferred to the best friend vector. More precisely, flags  $\cF$ and $\cF'$ in that example have best friend vectors $(2,4,8)$ an $(2,2,8)$, respectively. Moreover, the best friend vector might not even be an increasing sequence of divisors of $n$. 

\begin{example}\label{ex: seq is not a flag}
Consider any flag of type $(2, 5, 8)$ on $\bbF_{q^{16}}$ of the form
$$
\cF=(\bbF_{q^2}, \cU, \bbF_{q^8}).
$$
Since $\gcd(5, 16)=1$, the best friend of $\cU$ is $\bbF_q$ and the best friend vector of $\cF$, which is $(2,1,8)$, is not an increasing sequence.
\end{example}

However, in the previous example, the sequence of best friends was $\bbF_{q^2}, \bbF_{q}, \bbF_{q^8}$ that, up to order, constitutes a sequence of nested subfields. Consequently, the best friend of the flag $\cF$ still coincides with the best friend of one of its subspaces. As we can see in the following example, this property is also not true in general.

\begin{example}\label{ex: BF (4,3)}
Consider any element $\alpha\in\bbF_{q^{24}}\setminus \bbF_{q^{12}}$ and the subspace $\cU= \bbF_{q^{12}}\oplus \bbF_{q^{3}}\alpha,$ which has dimension $15$ over $\bbF_{q}$. Clearly $\bbF_{q^3}$ is a friend of $\cU$ and, since $\gcd(15, 24)=3$, it is its best friend. Take now the flag $\cF=(\bbF_{q^4}, \cU)$ of type $(4, 15)$ on $\bbF_{q^{24}}$. Its best friend vector is clearly $(4,3)$. As a result, the best friend of $\cF$ is the ground field $\bbF_q$.
\end{example}

As showed in Example \ref{ex: BF (4,3)}, consecutive subspaces in a flag can have non-nested best friends and hence, as we can see, the best friend of a flag does not need to coincide with the best friend of any of its subspaces. In the next section we study how the presence of consecutive subfields of $\bbF_{q^n}$ in the sequence of best friends of a flag $\cF$ determines a set of possibilities for the minimum distance and for the type vector of $\orb(\cF)$. This study is undertaken by considering all the different options for two subfields in the sequence of best friends of a flag: equal, different but nested or not nested.

\section{Parameters interdependence}\label{sec:interplay} 
It is clear that the best friend vector of $\orb(\cF)$ completely determines the best friend of $\orb(\cF)$ even though the converse is not true.
In this section we exhibit how the knowledge of the best friend vector of a flag $\cF$ provides more accurate information about the minimum distance of $\orb(\cF)$ and, at the same type, impose conditions on the type vector of $\cF$ itself. Concerning the cardinality of $\orb(\cF)$, Theorem \ref{th: bf and size flag}, it can be calculated directly from the best friend. Moreover, as showed in \cite{Cyclic}, we always have the next connection.

\begin{theorem}\label{th: bf vector and cardinalities}
	Let $\cF=(\cF_1, \dots, \cF_r)$ be a flag on $\bbF_{q^n}$. Then, for every $1\leq i\leq r$, the value $|\orb(\cF_i)|$ divides $|\orb(\cF)|$. More precisely, if the best friend vector of $\cF$ is $(m_1, \dots, m_r)$ and $m=\gcd(m_1, \dots, m_r)$, then $|\orb(\cF)|= |\orb(\cF_i)|\cdot\frac{q^{m_i}-1}{q^m-1},$ for every $1\leq i\leq r$.
\end{theorem}

This result comes from the relationship between the best friend and the stabilizer subgroup of a flag under the action of $\bbF_{q^n}^\ast.$ More precisely, if $\cF=(\cF_1,\ldots, \cF_r)$ is a flag on $\bbF_{q^n}$ with best friend $\bbF_{q^m}$ and such that the best friend of $\cF_i$ is  $\bbF_{q^{m_i}}$, then 
\begin{equation}\label{eq: stab flag and subspace}
\stab(\cF)=\bbF_{q^m}^\ast  \ \text{and} \ \ \stab(\cF_i)=\bbF_{q^{m_i}}^\ast, \ \text{for every} \ 1\leq i\leq r.
\end{equation}
As a consequence we can straightforwardly derive the following result:
\begin{proposition}
	Let $\cF=(\cF_1,\ldots, \cF_r)$ a flag with best friend vector $(m_1, \ldots, m_r)$. Take $m= \gcd(m_1, \ldots, m_r)$. Hence $|\orb(\cF)|=|\orb(\cF_i)|$ if, and only if, $m=m_i$. Otherwise  $|\orb(\cF)| > |\orb(\cF_i)|$.
\end{proposition}

We continue by analyzing how with the help of the best friend vector we can better estimate the minimum distance of a cyclic orbit flag code.

\subsection{Best friend vector and minimum distance}\label{subsec: bf and dist}

As pointed out before, in \cite{Cyclic}, the authors showed that the knowledge of the best friend of the generating flag $\cF$ can give some estimates about the minimum distance of the code $\orb(\cF)$. In that paper it was proved that, if $\cF, \cF'$ are flags with $\bbF_{q^m}$ as their best friend, then $2m$ divides $d_f(\cF, \cF')$. As a direct consequence, one has that
\begin{equation}\label{eq: bounds min dist BF}
2m\leq d_f(\orb(\cF)) \leq D^{(\bt, n)}.
\end{equation}

\noindent Let us see how the additional knowledge of the best friend vector of $\cF$ can improve considerably this lower bound for the minimum distance of $\orb(\cF)$.

\begin{theorem}\label{th: bound min dist seq BF}
Let $\cF=(\cF_1, \dots, \cF_r)$ be a flag on $\bbF_{q^n}$ with best friend vector $(m_1, \dots, m_r)$. Then it holds
$$
d_f(\orb(\cF))\geq 2\min\{m_i \ | \ 1\leq i\leq r\}.
$$
\end{theorem}
\begin{proof}
Take any $\alpha\in\bbF_{q^n}^\ast\setminus \stab(\cF)$ and compute $d_f(\cF, \cF\alpha).$ Let us write $m_j=\min\{m_i \ | \ 1\leq i\leq r\}$. If $\alpha \notin\bbF_{q^{m_{j}}}=\stab(\cF_j)$, then we have $\cF_j\neq\cF_j\alpha$ and
$$
d_f(\cF, \cF\alpha) \geq d_S(\cF_j, \cF_j\alpha)\geq 2m_j.
$$
On the other hand, if $\alpha \in\bbF_{q^{m_{j}}}=\stab(\cF_j)$, since $\alpha\notin\stab(\cF)$, there exists at least a subspace $\cF_i$ in $\cF$ such that $\cF_i\neq\cF_i\alpha$. In this case, it clearly holds
$$
d_f(\cF, \cF\alpha) \geq d_S(\cF_i, \cF_i\alpha)\geq 2m_i \geq 2m_j.
$$
Hence, we conclude that $d_f(\orb(\cF))\geq 2m_j= 2\min\{m_i \ | \ 1\leq i\leq r\}$.
\end{proof}

\begin{remark}
Notice that Theorem \ref{th: bound min dist seq BF} notably improves the lower bound given in (\ref{eq: bounds min dist BF}). If $\cF$ is a flag with best friend $\bbF_{q^m}$ and best friend vector $(m_1, \dots, m_r)$, by means of Proposition \ref{prop: bf intersection bfs}, we have
$$
m=\gcd(m_1, \dots, m_r) \leq \min\{ m_i \ | \ 1\leq i \leq r\}.
$$
Equality holds if, and only if, there is one index $i \in \{i_1, \ldots, i_r\}$ such that $m_i$ divides the rest ones. For this particular $m_i$, it holds $m_i=m$.
\end{remark}

 In view of this result, we can appreciate that the best friend vector allows us to better estimate the minimum distance than just the best friend. Hence,  it is worth asking what features of this new invariant may be helpful in this direction. For instance, we may ask if the number of subspaces in $\cF$ having the same best friend $\bbF_{q^m}$ than $\cF$ could have any relevance in order to bounding the minimum distance of $\orb(\cF)$. This approach was already suggested in \cite{GenGalois} where this number was taken into account to propose lower bounds for the minimum distance. Here, we complete this idea  by also considering those subspaces having as best friend a subfield bigger than $\bbF_{q^m}$. In this way we can also provide upper bounds for the minimum distance improving that in (\ref{eq: bounds min dist BF}).

\begin{theorem}\label{th:number subspaces BF}
	Let $\cF$ be a flag on $\bbF_{q^n}$ with $(m_1,\ldots, m_r)$ as best friend vector. Consider $m=\gcd(m_1,\ldots, m_r)$ and $j=|\{i \ | \ m_i=m \}|$. Then we have:
	\begin{enumerate}
		\item If $j > 0$, then $d_f(\orb(\cF))\geq 2mj.$  In case  $j=0$, it holds $d_f(\orb(\cF))> 2m$. Conversely, if $d_f(\orb(\cF))=2m$, then $j=1$.
		\item Assume that $j < r$. Let $1\leq i\leq r$ be any index such that $m_i \neq m$, then  
		$$
		2mj \leq d_f(\orb(\cF))\leq  D^{(\bt, n)}(i).
		$$	
	\end{enumerate}
\end{theorem}
\begin{proof}
 Consider  the value $j=|\{i \ | \ m_i=m\}|$.
For $(1)$, we start assuming $j>0$ and we take $1\leq i_1 < \dots < i_j\leq r$ such that $m_{i_k}=m$. If $\alpha\notin \stab(\cF)=\bbF_{q^m}^\ast = \stab(\cF_{i_k})$, then we have $\cF_{i_k}\neq\cF_{i_k}\alpha.$ Consequently,
$$
d_f(\cF, \cF\alpha)\geq \sum_{k=1}^j d_S(\cF_{i_k}, \cF_{i_k}\alpha) \geq \sum_{k=1}^j 2m = 2mj,
$$
where the last inequality follows from Theorem \ref{th: bf and dist}. As a result, $d_f(\orb(\cF))\geq 2mj,$ as stated. 
On the other hand, if $j=0$, by means of Theorem \ref{th: bound min dist seq BF}, we have $d_f(\orb(\cF))\geq 2\min\{m_i \ | \ 1\leq i\leq r\} > 2m.$
In particular, if $d_f(\orb(\cF))=2m$, then $j\neq 0$ and $j< 2$, i.e., $j=1$.

To prove $(2)$, suppose that, for some $1\leq i\leq r$, the subspace $\cF_i$ has best friend $\bbF_{q^{m_i}}\neq \bbF_{q^m}.$ By Proposition \ref{prop: bf intersection bfs}, we clearly have $\bbF_{q^m}\subsetneq \bbF_{q^{m_i}}$ and we can find elements in $\alpha\in\bbF_{q^{m_i}}\setminus\bbF_{q^m}.$ Recall that $\stab(\cF)=\bbF_{q^m}^\ast$ and $\stab(\cF_i)=\bbF_{q^{m_i}}^\ast$ (see (\ref{eq: stab flag and subspace})), then we have $\cF\neq \cF\alpha$ whereas $\cF_i=\cF_i\alpha$. Finally,
$$
d_f(\orb(\cF))\leq d_f(\cF, \cF\alpha) \leq D^{(\bt, n)}(i).
$$
Repeating this argument for every subspace with best friend different from $\bbF_{q^m}$ gives the stated bound.
\end{proof}

The following example reflects that the converse of this result does not hold.
\begin{example}
	Consider the flag
$$\cF=(\bbF_{q^4}, \bbF_{q^{12}}, \bbF_{q^{12}}\oplus \bbF_{q^3}\alpha),$$
for some $\alpha\in\bbF_{q^{24}}\setminus\bbF_{q^{12}}.$ Its type vector is $(4,12,15)$ and it has best friend vector $(m_1, m_2, m_3)=(4,12,3).$ Clearly, the best friend of the flag is $\bbF_q,$ since $m=\gcd(4,12,3)=1.$ By means of Theorem \ref{th: bound min dist seq BF}, we know that
$$
d_f(\orb(\cF))\geq 2\cdot 3 = 2\cdot 1 \cdot 3. 
$$
However, we have $j=|\{ i  \ | \ m_i=m\}|=0\neq 3.$ 

On the other hand, take the same $\alpha\in\bbF_{q^{24}}\setminus\bbF_{q^{12}}$ and form a a flag 
$$
\cF'=(\bbF_{q^{4}}, \cF'_2, \bbF_{q^{12}}, \bbF_{q^{12}}\oplus \bbF_{q^3}\alpha)
$$
of type $\bt=(4,5,12,15)$ on $\bbF_{q^{24}}.$ In this case, the best friend vector is $(m_1, \dots, m_4)=(4,1,12,3)$ and $m=m_2=1.$ Notice that, since $m_3=12\neq m=1,$ by Theorem \ref{th:number subspaces BF}, we have
$$
d_f(\orb(\cF'))\leq D^{(\bt, 24)}(3)=8+10+0+6=24.
$$
Hence, it also holds $d_f(\orb(\cF')) < D^{(\bt, 24)}(2)=2+0+14+18=34$. However, we have $m=m_2=1.$
\end{example}

 We can go further and consider the case where a subfield other than the best friend of a flag $\cF$ is, in turn, the best friend of several of its subspaces.  Paying attention to this fact permits us to directly improve the previous upper bound as follows. 
 
\begin{theorem}\label{cor: bound dist diff bf several times}
Let $\cF=(\cF_1, \dots, \cF_r)$ be a flag on $\bbF_{q^n}$ with best friend vector $(m_1, \ldots, m_r)$. Let $l$ be a positive integer with $l \neq \gcd(m_1, \ldots, m_r)$ such that $l=m_i$ for exactly $1\leq s < r$ entries in $(m_1, \ldots, m_r)$. Then
$$
d_f(\orb(\cF)) \leq  D^{(\bt, n)}(i_1, \dots, i_s).
$$
\end{theorem}

\begin{remark}
Observe that
$
D^{(\bt, n)}(i_1, \dots, i_s) < D^{(\bt, n)}(i_j), \ \forall j=1, \dots, s,
$
which makes the last bound be tighter than the one in Theorem  \ref{th:number subspaces BF}. On the other hand, a subfield $\bbF_{q^l}\neq \bbF_{q^m}$ can appear in the sequence of best friends of a flag $\cF$ at most $r-1$ times. Otherwise, $\bbF_{q^l}$ would be the best friend of all the subspaces in $\cF$ and, by means of Proposition \ref{prop: bf intersection bfs}, also its best friend. 
\end{remark}

Another property of the best friend vector of  flag $\cF$ that we can also take into account is that, even if its entries are not equal to $m$, some of them can give a (possibly unordered) sequence of consecutive divisors, that is, the corresponding best friends of the subspaces of $\cF$ might be nested. This situation is considered in the next result.

\begin{theorem}\label{theo: neste seq bf vector}
Let $\cF=(\cF_1, \dots, \cF_r)$ be a flag on $\bbF_{q^n}$ with best friend vector $(m_1, \ldots, m_r)$ and put $m=\gcd(m_1,\ldots,m_r)$. Assume that  $(m_1, \ldots, m_r)$ contains  $1\leq s < r$ (possibly unordered) entries $m_{i_1}, \dots, m_{i_s}$  different from $m$ and such that $m_{i_k}$ divides $m_{i_{k+1}}$, for every $1 \leq k <s$.
Then
$$
d_f(\orb(\cF))\leq D^{(\bt, n)}(i_1,\dots, i_s).
$$
\end{theorem}

\begin{proof}
By hypothesis, we know that $\bbF_{q^{m_{i_1}}} \subseteq  \bbF_{q^{m_{i_2}}} \subseteq \dots \subseteq \bbF_{q^{m_{i_s}}}$ are subfields different from $\bbF_{q^m}$ appearing in the sequence of best friends of the subspaces in $\cF$, possibly not in this order. Since $m\neq m_{i_1}$,  we can find elements $\alpha\in\bbF_{q^{m_{i_1}}}^\ast\setminus \bbF_{q^m}^\ast$. Recall that $\stab(\cF)=\bbF_{q^m}^\ast$ and, for every $1\leq i\leq r$, it also holds $\stab(\cF_i)=\bbF_{q^{m_i}}^\ast$. Thus, the flag $\cF\alpha$ is different from $\cF$ but $\cF_{i_k}=\cF_{i_k}\alpha$ for every $1\leq k \leq s$. Then the distance vector associated to the pair of flags $\cF$ and $\cF\alpha$ contains zeros in the (possibly not ordered) positions $i_1, \dots, i_s$, which leads to
$$
d_f(\orb(\cF))\leq d_f(\cF, \cF\alpha) \leq  D^{(\bt, n)}(i_1,\dots, i_s).
$$
\end{proof}

Following with Example \ref{ex:flags different best friend vector}, we can see that flag codes with the same best friend can have different parameters if they do not share their best friend vector.

\begin{example}
Consider integers $s\geq 2$ and $n=8s$ and take the flags $\cF=(\bbF_{q^2}, \bbF_{q^4}, \bbF_{q^8})$ and  $\cF'=(\bbF_{q^2}, \bbF_{q^2}\oplus \bbF_{q^2}\beta, \bbF_{q^8})$ on $\bbF_{q^n}$ given in Example \ref{ex:flags different best friend vector}. These two flags have best friend $\bbF_{q^2}$ and respective best friend vectors $(2,4,8)$ and $(2,2,8)$. Clearly both codes $\orb(\cF)$ and $\orb(\cF')$ have the same cardinality $\frac{q^n-1}{q^2-1}$. However, the list of sizes of their projected codes differ. More precisely, as stated in Theorem \ref{th: bf vector and cardinalities}, it holds
$$
|\orb(\bbF_{q^4})|=\frac{q^n-1}{q^4-1} \neq \frac{q^n-1}{q^2-1} = |\orb(\bbF_{q^2}\oplus \bbF_{q^2}\beta)|.
$$
Concerning the mininimum distance, by application of Theorem \ref{th:number subspaces BF} and Theorem \ref{theo: neste seq bf vector}, we have
$$
4= 2\cdot 2\cdot 1\leq d_f(\orb(\cF))\leq D^{((2,4,8),n)}(2,3)=4
$$
and then $d_f(\orb(\cF))=4$. On the other hand, since the value $2$ appears twice in the best friend vector of $\cF'$,  by means of Theorem \ref{th:number subspaces BF}, we get
$$
d_f(\orb(\cF')) \geq 2\cdot 2\cdot 2=8.
$$
\end{example}

The next example exhibits how the knowledge of the best friend vector considerably improves the estimates for the minimum distance of cyclic orbit flag codes, compared to the bounds obtained in Proposition \ref{prop: distance bounds BF}, where just the best friend of the flag (and not the ones of its subspaces) is taken into consideration. 

\begin{example}
Consider a flag $\cF$ of type $\bt=(2,4,5,12,15,18,21)$ on $\bbF_{q^{24}}$ with best friend vector $(m_1, \dots, m_7)=(2,4,1,12,3,3,3)$. The best friend of $\cF$ is the ground field $\bbF_q$. In this case, since $j=|\{ i \ | \ m_i=m=1\}|=1$, the lower bound given in Proposition \ref{prop: distance bounds BF} and  Theorem \ref{th: bound min dist seq BF} coincide and $d_f(\orb(\cF))\geq 2$. Concerning upper bounds, according to Proposition \ref{prop: distance bounds BF}, that is, just looking at the best friend of $\cF$, the minimum distance of $\orb(\cF)$ satisfies 
$$
d_f(\orb(\cF))\leq D^{(\bt, 24)}= 4+8+ 10+ 24 +18+12+6= 82.
$$
On the other hand, if take into consideration the best friend vector $(2,4,1,12,3,3,3)$, we observe that the previous bound can be considerably improved:
\begin{itemize}
\item Since $m_5=m_6=m_7=3\neq m=1$, by Theorem \ref{cor: bound dist diff bf several times}, we obtain 
$$
d_f(\orb(\cF))\leq D^{(\bt, 24)}(5,6,7)= 4+8+10+6+0+0+0=28.
$$
\item From the subsequence of divisors $m_1=2$, $m_2=4$ and $m_4=12$  and by Theorem \ref{theo: neste seq bf vector}, we get
$$
d_f(\orb(\cF))\leq D^{(\bt, 24)}(1,2,4)= 0+0+2+0+6+12+6= 26.
$$
\item The same result, but considering the subsequence of divisors given by $m_5=m_6=m_7=3$ and $m_4=12$ leads to
$$
d_f(\orb(\cF))\leq D^{(\bt, 24)}(4,5,6,7)=4+8+10+0+0+0+0=22. 
$$
\end{itemize}
\end{example}

\subsection{Best friend vector and type vector}\label{subsec: bf and type}

 In this part, we analyze how the knowledge of the best friend vector of a flag $\cF$ on $\bbF_{q^n}$, hence the one of $\orb(\cF)$, conditions its type vector and even the dimension of the ambient space. In order to simplify our approach, we work first with flags of length two on $\bbF_{q^n}$ with prescribed best friend vector $(m_1,m_2)$. We distinguish different possibilities based on whether or not the value $\gcd(m_1, m_2)$ belongs to $\{m_1, m_2\}$. The results obtained for this particular situation will give us the clue to address the case of flags of any length.

 \begin{theorem}\label{th: bf no inclusion}
Let $\cF=(\cF_1, \cF_2)$ be a flag of type $(t_1, t_2)$ on $\bbF_{q^n}$ with best friend vector $(m_1,m_2)$. 
Then $t_2$ is a multiple of $m_2$ and the following statements hold:
\begin{enumerate}
	\item If  $\gcd(m_1, m_2)=m_i$,  where $i \in\{1,2\}$, then  $t_2 \geq t_1 + m_i$.
	\item If  $\gcd(m_1, m_2)\notin \{m_1,m_2\}$, then  $t_2\geq t_1 + \max\{m_1, m_2\}$.
\end{enumerate}
\end{theorem}

\begin{proof}
Let us prove $(1)$. If $m_1=m_2=m$, the result follows from the fact that $m$ divides both $t_1$ and $t_2$, and $t_1 < t_2$. Let us assume that $m_1 \neq m_2$ but $\gcd(m_1, m_2)=m_1$, that is,  $\bbF_{q^{m_1}}\subsetneq\bbF_{q^{m_2}}$. We can consider some element $\alpha\in\bbF_{q^{m_2}}\setminus\bbF_{q^{m_1}}$. Hence, $\cF_1\neq\cF_1\alpha$ but $\cF_2=\cF_2\alpha$. In this case,
$$
\bd(\cF, \cF\alpha)=(d_S(\cF_1, \cF_1\alpha), 0).
$$
Moreover, by means of Theorem \ref{th: bf and dist}, we have that $2m_1$ divides $d_S(\cF_1, \cF_1\alpha)$ and, in particular, $d_S(\cF_1, \cF_1\alpha)\geq 2m_1$. Moreover, Theorem \ref{th: char dist vect} implies that
$$
2m_1 \leq d_S(\cF_1, \cF_1\alpha) = | 0 - d_S(\cF_1, \cF_1\alpha) | \leq 2(t_2 -t_1)
$$
and the first statement holds in case $\gcd(m_1, m_2)=m_1$. The case $\gcd(m_1, m_2)=m_2$ is analogous.

To prove $(2)$, suppose that  $\gcd(m_1, m_2) \notin \{m_1,m_2\}$, that is,  $\bbF_{q^{m_1}}\cap\bbF_{q^{m_2}} \notin\{ \bbF_{q^{m_1}}, \bbF_{q^{m_2}} \}$. In such a case, we can find elements $\alpha \in\bbF_{q^{m_1}}\setminus\bbF_{q^{m_2}}$ and $\beta \in\bbF_{q^{m_2}}\setminus\bbF_{q^{m_1}}$ and
$$
\bd(\cF, \cF\alpha)= (0, d_S(\cF_2, \cF_2\alpha)), \ \ \bd(\cF, \cF\beta)= (d_S(\cF_1, \cF_1\beta), 0).
$$
Moreover, we have $d_S(\cF_2, \cF_2\alpha)\geq 2m_2$ and $d_S(\cF_1, \cF_1\beta)\geq 2m_1$. Hence, by means of Theorem \ref{th: char dist vect}, we obtain
$$
\begin{array}{rccccl}
	2m_1 & \leq & d_S(\cF_1, \cF_1\beta)  & = & | 0 - d_S(\cF_1, \cF_1\alpha) | & \leq 2(t_2 -t_1)\\
	2m_2 & \leq & d_S(\cF_2, \cF_2\alpha) & = & | d_S(\cF_2, \cF_2\alpha) - 0 | & \leq 2(t_2 -t_1)\\
\end{array}
$$
and conclude that $t_2\geq t_1+m_1$ and  $t_2\geq t_1+m_2$, which finishes the proof.
\end{proof}

\begin{example}
For  a flag $\cF$ in $\bbF_q^n$ with $n=24$ and  best friend vector $(4,3)$, in light of Theorem \ref{th: bf no inclusion}, the type vectors $(4,6)$, $(8, 9)$, $(12, 15)$, $(16, 18)$ and $(20, 21)$ are not allowed. 
\end{example}

The previous result can be iteratively applied in order to determine bounds for the dimensions in the type vector of a flag of any length, provided its best friend vector.
\begin{corollary}
Let $\cF$ be a flag of type $(t_1, \dots, t_r)$ on $\bbF_{q^n}$ with best friend vector $(m_1, \dots, m_r)$. For every $1\leq i < r$, we have
$$
t_{i+1} \geq 
\left\lbrace
\begin{array}{ll}
t_i + m_i  & \text{if} \ \gcd(m_i, m_{i+1})= m_i,\\
t_i + m_{i+1} & \text{if} \ \gcd(m_i, m_{i+1})= m_{i+1},\\
t_i + \max\{m_i, m_{i+1}\} & \text{otherwise}. \\
\end{array}
\right.
$$
\end{corollary}

Notice that, even if the previous results give information on the type vector in terms of the best friend vector, their proofs are based on distance vectors properties. The following results provide complementary bounds for the dimensions in the type vector of a flag. In this case, we use the nested structure of flags, combined with the properties of towers of subfields of $\bbF_{q^n}.$ To this end, we come back to flags of length two and then extract conclusions for the general case.

\begin{lemma}\label{lemma: bf and containing field}
Consider subfields $\bbF_{q^{m_1}}$ and $\bbF_{q^{m_2}}$ of $\bbF_{q^n}$ and let $\cU$ be an $\bbF_{q^{m_2}}$-subspace of $\bbF_{q^n}.$ If $\bbF_{q^{m_1}}\subseteq \cU$, then $\cU$ also contains the minimum field containing both $\bbF_{q^{m_1}}$ and $\bbF_{q^{m_2}},$ that is, the subfield $\bbF_{q^l}$, with $l=\lcm(m_1, m_2).$ 
\end{lemma}
\begin{proof}
Consider $\bbF_q$-basis $\{1, \alpha, \dots, \alpha^{m_1-1}\}$ and $\{1, \beta, \dots,  \beta^{m_2-1}\}$ of $\bbF_{q^{m_1}}$ and $\bbF_{q^{m_2}}$, respectively. Since scalar multiplication by elements in $\bbF_{q^{m_2}}$ is closed in $\cU$, then it clearly contains the set
$$
\{ \alpha^i \beta^j \ | \ 0\leq i\leq m_1-1,  \ 0\leq j\leq m_2-1\},
$$
which is an $\bbF_q$-basis of the minimum field containing both $\bbF_{q^{m_1}}$ and $\bbF_{q^{m_2}}$, that is, $\bbF_{q^l}$ with $l=\lcm(m_1, m_2).$
\end{proof}

The previous result has clear consequences for the type vector of flags as we can see in the next result.

\begin{theorem}\label{th: bound t_2 using subfields}
Consider a flag $\cF=(\cF_1, \cF_2)$ of type $(t_1, t_2)$ on $\bbF_{q^n}$ with best friend vector $(m_1, m_2)$.
Then $t_2$ is a multiple of $m_2$ satisfying $t_2\geq \lcm(m_1, m_2).$ Moreover, if $m_1$ does not divide $m_2$, then $t_2 \geq \lcm(m_1, m_2)+m_2$.
\end{theorem}

\begin{proof}
First of all, consider a flag $\cF'=\cF\alpha^{-1}$ for some $\alpha\in\cF_1\subset \bbF_{q^n}^\ast$. Notice that $1\in\cF'_1$ and both flags $\cF$ and $\cF'$ have the same type $(t_1, t_2)$ and best friend vector $(m_1, m_2)$. Hence $\bbF_{q^{m_2}}$ is the best friend of $\cF'_2$, and it is an $\bbF_{q^{m_2}}$-vector space. Thus, the value $m_2$ clearly divides $t_2$. Moreover, since $1\in\cF'_1\subset \cF'_2$, we have $\bbF_{q^{m_1}}\subseteq \cF'_1 \subset\cF'_2$. Hence, by means of Lemma \ref{lemma: bf and containing field}, $\cF'_2$ contains the subfield $\bbF_{q^{l}}$ with $l=\lcm(m_1, m_2)$ and then $t_2 \geq l$. Moreover, if $m_1$ does not divide $m_2$, we still have $t_2 \geq l= \lcm(m_1, m_2)\neq m_2$ and, if the equality holds, then $\bbF_{q^{m_2}}\subsetneq \bbF_{q^{l}} = \cF'_2$. In this case, the best friend vector of $\cF'$ would be $(m_1, l) \neq (m_1, m_2).$ As a consequence, $t_2$ is, at least, the next multiple of $m_2$, i.e., $t_2 \geq l+m_2 = \lcm(m_1, m_2)+m_2$, as stated.
\end{proof}

\begin{remark}
 In case $m_1$ divides $m_2$, the previous bound just says $t_2\geq l= \lcm(m_1, m_2)=m_2, $ which is a direct consequence of having the subfield $\bbF_{q^{m_2}}$ as a best friend. In this situation, the equality can hold; it suffices to consider the Galois flag of type $(m_1, m_2)$. The bound in case $m_1$ does not divide $m_2$ is also tight in some cases, as we can see in Example \ref{ex: BF (4,3)}: for $m_1=4$ and $m_2=3$, the dimension $t_2$ is $t_2=\lcm(4,3)+3=15$. On the other hand, and as stated in Example \ref{ex: not admissible types}, not every type vector is admissible. Notice that the second case also contemplates the situation in which $m_2$ divides $m_1$. For instance, if $\{1, \alpha\}$ is an $\bbF_{q^4}$-basis of $\bbF_{q^8}$, it suffices to consider  the flag $\cF=(\bbF_{q^4},\bbF_{q^4}\oplus\bbF_{q^2}\alpha)$ of type $(4, 6)$ on $\bbF_{q^8}$. For this flag, it holds: $m_1=4$, $m_2=2$ and $t_2=6= \lcm(4, 2)+ 2$.
\end{remark}

As stated before, Theorem \ref{th: bf no inclusion} and Theorem \ref{th: bound t_2 using subfields} provide different and complementary lower bounds for the dimension $t_2$ in the type vector of a flag of $\cF=(\cF_1, \cF_2)$. The next example shows that, in some cases, the bounds obtained in Theorem \ref{th: bf no inclusion} are better than the ones in Theorem \ref{th: bound t_2 using subfields} and vice versa.
	
\begin{example}
Take a type vector $(t_1, t_2)$ on $\bbF_{q^{24}}$ and fix the best friend vector $(4,3)$. We consider two cases:
\begin{itemize}
\item If $t_1=4,$ since $\gcd(4,3)=1\notin\{4,3\},$ by means of Theorem \ref{th: bf no inclusion}, we conclude that $t_2$ must be a multiple of $3$ with $t_2\geq t_1+\max\{4,3\}=8.$ In other words, we obtain $t_2\geq 9.$ On the other hand, given that $4$ does not divides $3$, Theorem \ref{th: bound t_2 using subfields} leads to $t_2\geq \lcm(4, 3)+3=15,$ which is a better lower bound for $t_2$.
\item On the contrary, if $t_1=12$,  Theorem \ref{th: bf no inclusion} ensures that $t_2\geq \lcm(4,3)+3=15$. On the other hand, by application of Theorem \ref{th: bound t_2 using subfields}, the dimension $t_2$ must be a multiple of $m_2=3$ satisfying $t_2\geq t_1+\max\{4,3\}=16,$ i.e., $t_2\geq 18.$
\end{itemize}
\end{example}

\begin{example}\label{ex: not admissible types}
Following with the parameters of Example \ref{ex: BF (4,3)}, and by means of Theorem \ref{th: bound t_2 using subfields}, we see that it is not possible to give a couple of nested subspaces with respective best friends $\bbF_{q^4}$ and $\bbF_{q^3}$ for every choice of the type vector. For instance, type vectors  $(4, 6)$, $(4, 9)$, $(4, 12)$ or $(8, 12)$ are not allowed.
\end{example}

From Theorem \ref{th: bound t_2 using subfields} we can derive the next result, which states that some combinations of subfields are not permitted as a part of the sequence of best friends of the subspaces of a flag (of any length, not necessarily two).

\begin{corollary}\label{cor: m1 and m2 not compatible}
Consider $m_1$ and $m_2$  divisors of $n$. If $\lcm(m_1, m_2)=n$, then there is no flag on $\bbF_{q^n}$ with both $m_1$ and $m_2$ as entries in its best friend vector. 
\end{corollary}
\begin{proof}
Consider a flag $\cF=(\cF_1, \ldots, \cF_r)$ on $\bbF_{q^n}$ and assume that there are different indices $1 \leq i_1, i_2 \leq r$, not necessarily ordered, such that  $\bbF_{q^{m_j}}$ is the best friend of $\cF_{i_j}$ for $j \in \{1,2\}$. In such a case, by means of Theorem \ref{th: bound t_2 using subfields}, the dimension of $\cF_{\max\{i_1, i_2\}}$ is equal to $\lcm(m_1, m_2)=n$, which is not possible according to Definition \ref{def: flag}.
\end{proof}

The next result, in turn, characterizes those values of $n$ that make it possible having two arbitrary fields in the sequence of best friends of a flag.
\begin{theorem}\label{th: sequence valid iff s geq 2}
Take positive integers $m_1$ and $m_2$. There are flags on $\bbF_{q^n}$ with best friend vector $(m_1, m_2)$ if, and only if, $n= s \cdot\lcm(m_1, m_2)$, for some integer 

$$
s\geq
\left\lbrace
\begin{array}{cl}
3 & \text{if} \ \ m_1=m_2, \\
2 & \text{otherwise.}
\end{array}
\right.
$$

\end{theorem}
\begin{proof}
Let us first assume that $\cF$ is a flag on $\bbF_{q^n}$ with $(m_1, m_2)$ as its best friend vector. In particular, both $\bbF_{q^{m_1}}$ and $\bbF_{q^{m_2}}$ are subfields of $\bbF_{q^n}$ and then $m_1$ and $m_2$ divide $n.$ As a consequence, the value $\lcm(m_1, m_2)$ also divides $n$ and we can write $n=s\cdot \lcm(m_1, m_2)$ for some positive integer $s$. From Corollary \ref{cor: m1 and m2 not compatible} we conclude that $s\neq 1$ and then $s\geq 2$. Moreover, if $m_1=m_2$, then $n=s\cdot\lcm(m_1, m_2)=sm_1$ and, by means of Theorem \ref{th: bf no inclusion}, it holds $t_2\geq t_1+m_1\geq 2m_1$. Consequently,  we have $n = sm_1 > t_2 \geq 2m_1,$ i.e., $s\geq 3$.

Conversely,  suppose that $m_1\neq m_2$ and $n= s \cdot \lcm(m_1, m_2)$ for some $s\geq 2$. Let us construct a flag $\cF=(\cF_1, \cF_2)$ with best friend vector $(m_1, m_2)$ as follows. If $m_1$ divides $m_2$, we just take the Galois flag $(\bbF_{q^{m_1}}, \bbF_{q^{m_2}})$. Otherwise, we consider $l=\lcm(m_1, m_2) > m_2 $ and the subfield $\bbF_{q^l}$ of $\bbF_{q^n}$. Note that the dimension of $\bbF_{q^n}$ as an $\bbF_{q^l}$-vector space is $s\geq 2$. Hence, if $\alpha\in\bbF_{q^n}\setminus\bbF_{q^l}$, the subspace $\bbF_{q^l}\oplus\bbF_{q^l}\alpha$ is a direct sum. In particular, the subspace $\cU=\bbF_{q^l}\oplus\bbF_{q^{m_2}}\alpha$ has dimension $l+m_2 < 2l \leq n$ and has $\bbF_{q^{m_2}}$ as a friend.
 Let us see that it is precisely the best friend of $\cU$. To do so, we consider any other friend $\bbF_{q^h}$ of $\cU$ and we prove that  $h\leq m_2$. Notice that $h$ divides $\dim(\cU)=l+m_2$, while $l$ does not. Thus $\bbF_{q^l}$ is not the best friend of $\cU$ and then there are elements in $\bbF_{q^l}$ not stabilizing $\cU$ (recall that $\stab(\cU)$ is the multiplicative group of the best friend of $\cU$). In particular, we can find an element $\beta\in\bbF_{q^l}^\ast\setminus\stab(\cU)$ and form the subspace
 $$
 \cU\beta= \bbF_{q^l} \oplus \bbF_{q^{m_2}}\alpha\beta,
 $$
which is also an $\bbF_{q^h}$-vector space. Now, if $\alpha\beta\in\cU$, then $\bbF_{q^{m_2}}\alpha\beta\subset \cU$ and $\cU=\cU\beta$, which is a contradiction. Hence 
$$
\cU+\cU\beta=\bbF_{q^l} \oplus \bbF_{q^{m_2}}\alpha\oplus\bbF_{q^{m_2}}\alpha\beta
$$
has dimension $l+2m_2\leq 2l\leq n$ and $\bbF_{q^h}$ is one of its friends. In particular, $h$ divides both $l+m_2$ and $l+2m_2$ and, consequently, it also divides $m_2$. We conclude that $\bbF_{q^{h}}\subset \bbF_{q^{m_2}}$, which proves that $\bbF_{q^{m_2}}$ is the best friend of $\cU$.
 
In case  $m_1=m_2$ and $n=sm_1$ with $s\geq 3$, we just need to consider a primitive element $\alpha$ of $\bbF_{q^n}$ and form the flag $\cF=(\cF_1, \cF_2)=(\bbF_{q^{m_1}}, \bbF_{q^{m_1}}\oplus \bbF_{q^{m_1}}\alpha)$ of type $(m_1, 2m_1)$ on $\bbF_{q^n}$. The best friend of $\cF_1$ is clearly $\bbF_{q^{m_1}}$. Now assume that $\bbF_{q^{h_2}}$ is a friend of $\cF_2$. In particular, $h_2$ divides $2m_1$. On the other hand, the subspace $\cF_2+\cF_2\alpha=\bbF_{q^{m_1}}\oplus \bbF_{q^{m_1}}\alpha \oplus \bbF_{q^{m_1}}\alpha^2$ is also a vector space over $\bbF_{q^{h_2}}$ and it has dimension $3m_1$. Hence, $h_2$ divides $3m_1$ as well and we conclude that $h_2$ divides $m_1$. This means that $\bbF_{q^{m_1}}$ is the best friend of $\cF_2$ and $\cF$ has best friend vector $(m_1, m_1)$.
\end{proof}

\begin{example}
The best friend vector $(4,3)$ is not valid on $\bbF_{q^{12}}$ since $\lcm(4,3)=12$ (see Corollary \ref{cor: m1 and m2 not compatible}). However, it is permitted for any value of $n=12s$, with $s\geq 2$ by means of Theorem \ref{th: sequence valid iff s geq 2}. In particular, for $n=24$, the flag $\cF$ given in Example \ref{ex: BF (4,3)} has $(4, 3)$ as best friend vector and it has been constructed following the ideas in the proof of Theorem \ref{th: sequence valid iff s geq 2}.
\end{example}

We finish this section by studying how the type vector of a flag $\cF$, then the type vector of the cyclic orbit code $\orb(\cF)$, is affected by the choice of the best friend vector of $\cF$ in case of considering flags of any length, not necessarily two. We do so by generalizing  Lemma \ref{lemma: bf and containing field} and Theorem \ref{th: bound t_2 using subfields}  but paying attention to the fact that, when the length of the flag is $r>2$, the entries $t_i$ of the type vector with $i \geq 3$ are influenced also by the entries $m_j$ in the best friend vector with $j \leq i$ due, once again, to the nested structure of flags. Let us first provide a lower bound for each dimension in the type vector in terms of the best friend vector. For the next result, we take a flag $\cF$ such that $1\in\cF_1$ (see Remark \ref{rem: 1 in F1}).

\begin{lemma}\label{th: bound type using r fields}
Let $\cF=(\cF_1, \dots, \cF_r)$ be a flag of type $(t_1, \dots, t_r)$ on $\bbF_{q^n}$ with $1\in\cF_1$ and  best friend vector $(m_1, \dots, m_r)$. Then, for every $1\leq i\leq r$, the subspace $\cF_i$ contains the subfield $\bbF_{q^{l_i}}$, with $l_i=\lcm(m_1, \dots, m_i)$.
\end{lemma}
\begin{proof}
We prove the result by induction on $1\leq i\leq r$.
For $i=1$, the result clearly holds. For $i=2$, it is proved in Lemma \ref{lemma: bf and containing field}.

Assume now that, for every $1< i \leq r$, we have that $\bbF_{q^{l_{i-1}}}\subseteq \cF_{i-1}$, with $l_{i-1}=\lcm(m_1, \dots, m_{i-1})$. Let us prove the result for $\cF_i$.
Notice that $\cF_i$ is a vector space over $\bbF_{q^{m_i}}$. By the induction hypothesis, it is satisfied that $\bbF_{q^{l_{i-1}}}\subseteq \cF_{i-1}\subset \cF_i$. Hence, by means of Lemma \ref{lemma: bf and containing field}, we conclude that $\bbF_{q^{l_{i}}}\subseteq \cF_i$, where 
$$
l_i=\lcm(l_{i-1}, m_i)=\lcm(\lcm(m_1, \dots, m_{i-1}), m_i)=\lcm(m_1, \dots, m_{i-1}, m_i),
$$
as stated.
\end{proof}

This result has a direct impact on the type vector configuration of a flag having a prescribed best friend vector.
\begin{corollary}\label{cor: first bounds t_i}
 Let $\cF=(\cF_1, \dots, \cF_r)$ be a flag of type $(t_1, \dots, t_r)$ on $\bbF_{q^n}$ and best friend vector $(m_1, \dots, m_r)$. Then, for every $1\leq i\leq r$, the dimension $t_i$ is a multiple of $m_i$ satisfying
 $$
 t_i\geq \lcm(m_1, \dots, m_i).
 $$
Equality holds if, and only if, $t_i= m_i= \lcm(m_1, \dots, m_i).$

\end{corollary}
\begin{proof}
Consider an element $\alpha\in\cF_1\subset \bbF_{q^n}^\ast$ and form the flag $\cF'=\cF\alpha^{-1}.$ This flag has both the same type and best friend vectors as $\cF$ and satisfies $1\in\cF'$. For every $1\leq i\leq r,$ we apply Theorem \ref{th: bound type using r fields} to the subspace $\cF'_i$ and conclude that $\bbF_{q^{l_i}}\subseteq \cF'_i$, with $l_i=\lcm(m_1, \dots, m_i)$. Hence $t_i\geq l_i=\lcm(m_1, \dots, m_i)$ and the equality is satisfied if, and only if, $\cF'_i=\bbF_{q^{l_i}}= \bbF_{q^{m_i}}$ but, since the best friend of $\cF'_i$ is precisely $\bbF_{q^{m_i}}$, it must hold $m_i=l_i.$
\end{proof}

Last, we apply this result in order to discard many best friend vectors on $\bbF_{q^n}.$
\begin{corollary}\label{cor: not possible seq bf}
Let $m_1, \dots, m_r$ be divisors of $n$. If $\lcm(m_1, \dots, m_r)=n,$ then there is no flag on $\bbF_{q^n}$ whose  best friend vector has $m_1, \dots, m_r$ as its entries.
\end{corollary}

The previous corollary leads to the following result, which states a necessary condition for a maximal subfield of $\bbF_{q^n}$ to be the best friend of a subspace of a flag.
\begin{corollary}\label{cor: maximality}
Let  $\bbF_{q^m}$ be a maximal subfield of $\bbF_{q^n}$. If $m$ is an entry in the best friend vector of a flag $\cF$, then the rest of components on it are divisors of $m$.
\end{corollary}
\begin{proof}
It suffices to consider a subfield $\bbF_{q^l}$ of $\bbF_{q^n}$ not being a subfield of $\bbF_{q^m}$. Observe that the minimum field containing both $\bbF_{q^m}$ and $\bbF_{q^l}$ is the whole $\bbF_{q^n}$ by maximality of $\bbF_{q^m}.$ Hence, $\lcm(m, l)=n$ and Corollary \ref{cor: not possible seq bf} concludes the proof.
\end{proof}

\begin{example}
On $\bbF_{q^{12}}$, there is no flag $\cF=(\cF_1, \cF_2)$ with best friend vector $(4,3)$. In fact, with this best friend vector, and according to Theorem \ref{th: bound t_2 using subfields}, the second dimension $t_2$ must satisfy $t_2\geq 15 > 12$. This is due to the maximality of $\bbF_{q^{4}}$ as a subfield of $\bbF_{q^{12}}.$ On the other hand, if $n=24$, the same best friend vector is allowed (see Example \ref{ex: BF (4,3)}). Similarly, due to the maximality of $\bbF_{q^8}$ as a subfield of $\bbF_{q^{24}}$, and as a consequence of Corollary \ref{cor: maximality}, if $8$ is a component in the best friend vector of a flag, the rest of components in that vector are forced to belong to $\{1,2,4,8\}$. Likewise, if $\bbF_{q^{12}}$ is the best friend of a subspace of a flag on $\bbF_{q^{24}}$, then $\bbF_{q^8}$ is not permitted as the best friend for other subspaces in the same flag.
\end{example}

We end the section with a partial generalization of Theorem \ref{th: sequence valid iff s geq 2}. There, we proved the existence of flags with best friend vector $(m_1, m_2)$ on $\bbF_{q^n}$ if, and only if, $n=s\cdot\lcm(m_1, m_2)$ and $s\in\{2,3\}$. In this case, we give a sufficient condition on $n$ to ensure the existence of flags on $\bbF_{q^n}$ with best friend vector $(m_1, \dots, m_r)$ and propose a systematic construction of them. However, as we will see later with several examples, this condition is not always necessary.

\begin{theorem}\label{th: condition n existence flag best friend vector prescribed}
	Take $m_1, \ldots, m_r$ positive integers. For each index $2 \leq i \leq r$, define
	$$k_i=|\{j \in \{1, \ldots, i-1\}; \ \lcm(m_j, m_{j+1})= m_{j+1} > m_j \}|.$$
	Denote $k=k_r$ and $l=\lcm(m_1, \dots, m_r)$ and consider the value 
	$$s=\left\{ \begin{array}{ll}
		r-k & \text{if} \  m_r\neq l\\
		r-k+1 & \text{otherwise.}
		\end{array}\right.$$
 If $n \geq s \cdot \lcm(m_1, \ldots, m_r)$, then there exist flags $\cF=(\cF_1, \ldots, \cF_r)$ on $\bbF_{q^n}$ such that  $(m_1, \dots, m_r)$ is the best friend vector of $\orb(\cF)$.
\end{theorem}
\begin{proof}
For each index $1 \leq i \leq r$, let us denote $l_i=\lcm(m_1, \ldots, m_i)$. Hence $l=l_r$. Note that $k \leq r-1$. If $k < r-1$, then $s \geq 2$. In case $k=r-1$, the sequence $l_1, \ldots, l_r=l$ is strictly increasing. In particular, $l=m_r$ and, by definition, we have $s=r-k+1$, thus $s\geq 2$.

Assume that $n \geq sl$.  As $s\geq 2$, there is $\alpha \in \bbF_{q^n} \setminus \bbF_{q^l}$ such that $\bbF_{q^n}= \bbF_{q^l}(\alpha)$ and $\{1, \alpha, \ldots, \alpha^{s-1}\}$ is a basis of $\bbF_{q^n}$ over  $\bbF_{q^l}$. In particular $\bbF_{q^l} \oplus \bbF_{q^l}\alpha \oplus \cdots \oplus \bbF_{q^l}\alpha^{s-1}$ is a direct sum. Let us build the subspaces of $\cF$ by considering $\cF_1, \cF_2$ separately to adapt the process described in Theorem \ref{th: sequence valid iff s geq 2} to length $r>2$.
\begin{itemize}
	\item Construction of $\cF_1$. Just take $\cF_1=\bbF_{q^{m_1}}=\bbF_{q^{l_1}}$.
	\item Construction of $\cF_2$. We distinguish two cases:
	\begin{enumerate}
		\item If $\lcm(m_1, m_2)=m_2>m_1$,  take  $\cF_2=\bbF_{q^{l_2}}=\bbF_{q^{m_2}}$.
		\item If $m_1$ does not divide $m_2$  or  $m_1=m_2$ (then $l_2=m_2=m_1=l_1$), take  $\cF_2=\bbF_{q^{l_2}}\oplus \bbF_{q^{m_2}}\alpha$.
	\end{enumerate}

        \item   Construction of $\cF_i$. For $2 < i \leq r$, we take
$$
\cF_i = \bbF_{q^{l_i}} + \dots + \bbF_{q^{l_i}}\alpha^{i-k_i-2} +  \bbF_{q^{m_i}}\alpha^{i-k_i-1}.
$$

\end{itemize}
Notice that  for every $2\leq i < r$, we have $k_{i+1}\in\{ k_i, k_i +1\}$.  In particular, $k_{i+1}=k_i+1$ if $\lcm(m_i,m_{i+1}) = m_{i+1} > m_i$.  Hence, every $\cF_i$ consists of the sum of  $i-k_i$ summands.  In fact, in $\cF_{i+1}$ we have 
$$
i+1-k_{i+1}=  \left\lbrace
\begin{array}{lll}
i-k_i & \text{if} & k_{i+1}=k_i+1,\\
i-k_i+1  & \text{if} & k_i=k_{i+1}
\end{array}
\right.
$$
summands. In other words, every subspace $\cF_{i+1}$ is described as a sum having either the same number of summands as $\cF_i$ or exactly one more. Moreover, since $\bbF_{q^{m_i}}\subseteq \bbF_{q^{l_i}}\subseteq \bbF_{q^{l_{i+1}}}$, we have $\cF_{i} \subset \cF_{i+1}$ and we can form a flag $\cF=(\cF_1, \dots, \cF_r)$. 
Let us see that $\cF$ has best friend vector $(m_1, \dots, m_r)$, i.e., that $\bbF_{q^{m_i}}$ is the best friend of $\cF_i$, for every $1\leq i\leq r$. For $i=1$, the result clearly holds and the case $i=2$ has been already proved in Theorem \ref{th: sequence valid iff s geq 2}. For higher values of $i$, note first that $k_i\leq i-1$ and then $i-k_i-1\geq 0$. The case $i-k_i-1=0$ corresponds to the situation
$$
k_2=1, \dots, k_i=i-1,
$$
which happens if, and only if, every $m_j$ divides $m_{j+1}$ and $m_j\neq m_{j+1}$ for $1\leq j\leq i$. In such a case, we have $m_i=l_i$ and $\cF_i=\bbF_{q^{l_i}}+\bbF_{q^{m_i}}\alpha^0 = \bbF_{q^{l_i}}=\bbF_{q^{m_i}}$, which has dimension $t_i=l_i=m_i$  and best friend precisely $\bbF_{q^{m_i}}$.

Assume that $i-k_i-1>0$ and also that a subfield $\bbF_{q^{h_i}}$ of $\bbF_{q^n}$ is a friend of $\cF_i$. Let us see that  $h_i$ divides $m_i$. Observe that 
\begin{equation}\label{eq:i-k_i}
i-k_i\leq i+1-k_{i+1}\leq \dots \leq r-k_r =r-k \leq s.
\end{equation}

Hence, $i-k_i-1 \leq s-1$ and, since the elements $ 1, \alpha, \dots, \alpha^{s-1}$ are linearly independent over $\bbF_{q^{l}}$ and $\bbF_{q^{m_i}}\subset \bbF_{q^{l_i}}\subset\bbF_{q^l}$, then 
$$
\cF_i = \bbF_{q^{l_i}} \oplus \dots \oplus \bbF_{q^{l_i}}\alpha^{i-k_i-2} \oplus  \bbF_{q^{m_i}}\alpha^{i-k_i-1}
$$  
is a direct sum of dimension
$
t_i= l_i(i-k_i-1) + m_i.
$
Moreover, assuming that $\cF_i$ is a vector space over $\bbF_{q^{h_i}}$, implies that $h_i$ must divide $t_i$. Now we distinguish two situations:
\begin{itemize}
\item $i-k_i-1<s-1$. We consider the subspace $\cF_i\alpha$, which clearly is also a vector space over $\bbF_{q^{h_i}}$, and compute the sum $\cF_i+\cF_i\alpha$. As $i-k_i \leq s-1$, we get
	$$
	\cF_i+\cF_i\alpha =  \bbF_{q^{l_i}}\oplus\dots\oplus\bbF_{q^{l_i}}\alpha^{i-k_i-2}\oplus \bbF_{q^{l_i}}\alpha^{i-k_i-1}\oplus\bbF_{q^{m_i}}\alpha^{i-k_i}
	$$
	a direct sum  with $\dim(\cF_i+\cF_i\alpha)= l_i(i-k_i) + m_i$ and, since $\cF_i+\cF_i\alpha$ is a vector space over $\bbF_{q^{h_i}}$, then $h_i$ divides its dimension. Given that $h_i$ also divides $t_i=  l_i(i-k_i-1) + m_i$, we conclude that $h_i$ divides $\dim(\cF_i+\cF_i\alpha)-t_i= l_i$ and, as a consequence, it divides $m_i$ too. Hence, $\bbF_{q^{h_i}}\subseteq\bbF_{q^{m_i}}$ and $\bbF_{q^{m_i}}$ is the best friend of $\cF_i$.
	
\item $i-k_i-1=s-1$. From (\ref{eq:i-k_i}), this equality holds if, and only if, $s=r-k$ and $k_{j+1}=k_j+1$ for every $i\leq j< r$. This implies that $m_r\neq l$ and every $m_j$ divides $m_{j+1}\neq m_j,$ for $i\leq j < r$. Moreover, notice that $m_i\neq l_i$ (otherwise, $m_j=l_j$ for $i \leq j \leq r$ but we know that $m_r \neq l$). As a consequence, $\bbF_{q^{l_i}}$ is not the best friend of $\cF_i$ since $l_i$ does not divide $t_i= l_i(i-k_i-1)+m_i$ and hence $\stab(\cF_i)\neq \bbF_{q^{l_i}}^\ast$. We can find elements $\beta\in\bbF_{q^{l_i}}^\ast\setminus \stab(\cF_i)$ and compute $\cF_i\beta$. Taking into account that $\beta$ stabilizes $\bbF_{q^{l_i}}$ but it cannot stabilize $\bbF_{q^{m_i}}$, we have
	$$
	\cF_i\beta=\bbF_{q^{l_i}} + \dots + \bbF_{q^{l_i}}\alpha^{i-k_i-2} +  \bbF_{q^{m_i}}\alpha^{i-k_i-1}\beta.
	$$
	Notice that, if $\alpha^{i-k_i-1}\beta\in\cF_i$, since multiplication by elements in $\bbF_{q^{m_i}}$ is closed in $\cF_i,$ we would have $\bbF_{q^{m_i}}\alpha^{i-k_i-1}\beta\subset \cF_i$ and then $\cF_i=\cF_i\beta$, which is a contradiction. Thus $\bbF_{q^{m_i}}\alpha^{i-k_i-1}\beta\cap\cF_i=\{0\}$ and
	$$
	\cF_i + \cF_i\beta =\bbF_{q^{l_i}} \oplus \dots \oplus \bbF_{q^{l_i}}\alpha^{i-k_i-2} \oplus \bbF_{q^{m_i}}\alpha^{i-k_i-1}\oplus  \bbF_{q^{m_i}}\alpha^{i-k_i-1}\beta
	$$
is again a vector space over $\bbF_{q^{h_i}}$ and has dimension $l_i(i-k_i-1)+2m_i$. Hence $h_i$ divides both $t_i=l_i(i-k_i-1)+m_i$ and $l_i(i-k_i-1)+2m_i$. In particular, $h_i$ divides $m_i$ and $\bbF_{q^{m_i}}$ is the best friend of $\cF_i$.
\end{itemize}

Last, notice that $t_r=\dim(\cF_r)=l_r(r-k_r-1)+m_r= l(r-k-1)+m_r$. We consider two possibilities: 
\begin{itemize}
\item If $m_r=l$, we have $t_r= l(r-k) < l(r-k+1)= ls \leq n$.
\item Otherwise $m_r < l$ and it holds  $t_r = l(r-k-1) +m_r < l(r-k)= ls \leq n$.
\end{itemize}
We conclude that the chosen value of $n$ ensures the existence of flags $\cF$ on $\bbF_{q^{n}}$ with best friend vector $(m_1, \dots, m_r)$, that is, the cyclic orbit flag code $\orb(\cF)$ has also $(m_1, \dots, m_r)$ as best friend vector.
\end{proof}

\begin{remark}
Notice that, for $r=2$, we have
$$
k=k_2=\left\lbrace
\begin{array}{cl}
1  & \text{if} \  m_1 \  \text{divides} \ m_2 \ \text{and} \ m_1\neq m_2, \\ 
0  & \text{otherwise}.
\end{array}
\right.
$$
Hence, the value $s$ defined in Theorem \ref{th: condition n existence flag best friend vector prescribed} is
$$
s=\left\{ \begin{array}{ll}
		2 & \text{if} \  m_1 \ \text{does not divide} \ m_2, \\
		2 & \text{if} \  m_1 \ \text{divides} \ m_2 \ \text{and} \ m_1\neq m_2,\\
		3 & \text{if} \  m_1 = m_2.
		\end{array}\right.
$$
In other words, the choice of $s$ coincides with the one made in Theorem \ref{th: sequence valid iff s geq 2}.
\end{remark}

For some special choices of the best friend vector, the previous result is a characterization of the minimum value of $n$ needed for the existence of flags on $\bbF_{q^n}$ with the given best friend vector.

\begin{corollary}
Let $m$ be a positive integer and consider $r\geq 2$. There are flags with best friend vector $(m, \overset{(r)}{\ldots}, m)$ on $\bbF_{q^n}$ if, and only if, $n=sm$ with $s\geq r+1$.
\end{corollary}
\begin{proof}
Assume that $\cF$ is a flag of length $r$ on $\bbF_{q^n}$  with best friend vector $(m, \dots, m)$. Hence, $\cF$ has type $(t_1, \dots, t_r)$ and $m$ divides every $t_i$ and $n$. Put $n=sm$ and notice that $t_1 < \dots < t_r < n$. Hence, for every $1\leq i\leq r$, it holds $t_i \geq m i$. In particular, $n =sm > t_r\geq mr$ and then $s\geq r+1$, as stated.

For the converse, put $n= sm$ with $s\geq r+1$. We apply Theorem \ref{th: condition n existence flag best friend vector prescribed}, taking into account that $k=k_r=0$ and $m_r=m=l$ and the result holds. More precisely, the flag $\cF=(\cF_1, \dots, \cF_r)$ given in the proof of Theorem \ref{th: condition n existence flag best friend vector prescribed} is given by
$$
\cF_i= \bbF_{q^{m}}\oplus \bbF_{q^{m}}\alpha \oplus \dots \oplus \bbF_{q^{m}}\alpha^{i-1},
$$
for $1\leq i\leq r$, where $\{1, \alpha, \dots, \alpha^r, \dots, \alpha^{s-1}\}$ is an $\bbF_{q^m}$-basis of $\bbF_{q^n}$. 
\end{proof}

\begin{corollary}
Consider positive integers $m_1, \dots, m_r$ such that, for every $1\leq i< r$, the value $m_i$ divides $m_{i+1}$ and $m_i\neq m_{i+1}$. There are flags on $\bbF_{q^n}$ with best friend vector $(m_1, \dots, m_r)$ if, and only if, $n=sm_r,$ with $s\geq 2$.
\end{corollary}

\begin{proof}
Let $\cF$ be a flag  satisfying these properties, then $n=s\cdot\lcm(m_1, \dots, m_r)=sm_r$ and, by Corollary \ref{cor: not possible seq bf}, we know that $s\geq 2$.  For the converse, observe that, in this situation we have $k_i=i-1$ for every $1< i\leq r$. In particular, $k=k_r=r-1$. Moreover, $m_r=l=\lcm(m_1, \dots, m_r)$. Hence, if $n=sm_r$ with $s\geq r-k+1 = r-(r-1)+1= 2$, there are flags on $\bbF_{q^n}$ with best friend vector $(m_1, \dots, m_r)$ by application of Theorem \ref{th: condition n existence flag best friend vector prescribed}. More precisely, the flag constructed using the proof of such a result is the Galois flag $(\bbF_{q^{m_1}}, \dots, \bbF_{q^{m_r}})$ of type $(m_1, \dots, m_r)$ on $\bbF_{q^{n}}$ with $n=sm_r$ and $s\geq 2$.
\end{proof}

Despite for these particular cases the converse of Theorem \ref{th: condition n existence flag best friend vector prescribed} also holds, the following examples show that this is not true in general.

\begin{example}
Consider the best friend vector $(m_1, m_2, m_3)=(3,2,1)$ of length $r=3$. For this choice, we have $k_2=k_3=k=0$. Moreover, $l=\lcm(3,2,1)=6$ and $m_3\neq 6$. Hence, Theorem \ref{th: condition n existence flag best friend vector prescribed} ensures the existence of flags on $\bbF_q^{n}$ with the given best friend vector provided that $n\geq 3\cdot 6=18$. In fact, the flag proposed in the proof of that result is $\cF=(\cF_1, \cF_2, \cF_3)$ with subspaces
$$
\cF_1=\bbF_{q^3}, \quad \cF_2= \bbF_{q^6}\oplus\bbF_{q^2}\alpha, \quad \cF_2= \bbF_{q^6}\oplus\bbF_{q^6}\alpha\oplus \bbF_{q}\alpha^2,
$$ 
where $\{1, \alpha, \alpha^2\}$ are $\bbF_{q^6}$-linearly independent elements in $\bbF_{q^{6s}}$ and $s\geq 3$.

However, for this particular case, we can also construct a flag $\cF'$ on $\bbF_{q^{12}}$ with best friend vector $(3,2,1)$ as follows: take $\beta\in\bbF_{q^{12}}\setminus\bbF_{q^6}$ and $\gamma\in\bbF_{q^{12}}\setminus(\bbF_{q^6}\oplus\bbF_{q^2}\beta)$ and consider 
$$
\cF'=(\bbF_{q^3}, \ \bbF_{q^6}\oplus\bbF_{q^2}\beta, \ \bbF_{q^6}\oplus\bbF_{q^2}\beta\oplus\bbF_q\gamma)
$$
that clearly satisfies the required conditions.
\end{example}

\begin{example}
For the type vector $(m_1, \dots, m_5)=(2,4,8,1,1)$ of length $r=5$, we have $k_2=1$, $k_3=k_4=k_5=k=2$ and $m_5=1\neq l=\lcm(m_1, \dots, m_5)=8$. Hence, Theorem \ref{th: condition n existence flag best friend vector prescribed} guarantees the existence of flags on $\bbF_{q^n}$ whenever $n=8s$ and $s\geq r-k= 5-2=3$. However, for every choice of subspaces $\cU$ and $\cV$  of dimensions $\dim(\cU)=11$ and $\dim(\cV)=13$ satisfying $\bbF_{q^8}\subset \cU \subset\cV\subset\bbF_{q^{16}}$, the flag   
$$
\cF=(\bbF_{q^2}, \ \bbF_{q^4}, \ \bbF_{q^8}, \ \cU, \ \cV)
$$ 
has best friend vector $(2,4,8,1,1)$ and it is a flag on $\bbF_{q^{16}}$.
\end{example}

\section{Conclusions}\label{sec: conclusions}
 In this work we have introduced a new invariant for cyclic orbit flag codes: the best friend vector. This invariant depends exclusively on the generating flag $\cF$ and captures the way the best friend of $\orb(\cF)$ is obtained taking into account those of the subspaces of $\cF$. At the same time, it conditions the rest of parameters of $\orb(\cF)$ and provides more precise information about them than just the best friend. First of all, it permits to determine the cardinality of the orbit code as well as those of its projected codes. Moreover, paying attention to the configuration of the best friend vector we have derived better lower and upper bounds for the minimum distance. In particular, this study opens the door to find constructions of cyclic orbit flag codes having a prescribed value of the minimum distance, by taking into account the best friend vector of the generating flag as a crucial ingredient. 
On the other hand, we have also studied how this new invariant and the type vector of a flag are related.  Moreover, we have seen that not every best friend vector can be realized on $\bbF_{q^n}$. For flags of length $r=2$, we have completely determined the minimum value of $n$ making a best friend vector $(m_1, m_2)$ feasible on $\bbF_{q^n}$. On the other hand, for higher values of $r$, we have provided a sufficient condition on $n$ for flags on $\bbF_{q^n}$ with best friend vector $(m_1, \dots, m_r)$ to exist by exhibiting a systematic construction of such flags. For special choices of the best friend vector, we see that this condition on $n$ is also necessary. Nevertheless, determining the minimum value of $n$ for  which we can ensure the existence of flags with prescribed best friend vector on $\bbF_{q^n}$ is still an open question.

\end{document}